\documentclass[useAMS,usenatbib]{mn2e}
\usepackage{epsfig}
\usepackage{natbib}

\DeclareMathVersion{bold}

\title{The Mexican Hat Wavelet Family.\\ Application
to point source detection in CMB maps}

\author[Gonz\'alez-Nuevo et al.]{J. Gonz\'alez-Nuevo$^{1}$ \footnotemark,
F. Arg\"ueso$^{2}$, M. L\'opez-Caniego$^{3,4}$, L. Toffolatti$^{5}$,
\newauthor J. L. Sanz$^{3}$, P. Vielva$^{3}$, D. Herranz$^{3}$
\\  $^{1}$ SISSA-I.S.A.S, via Beirut 4, I-34014 Trieste, Italy
\\  $^{2}$ Departamento de Matem\'aticas, Universidad de Oviedo, Avda. Calvo Sotelo s/n, 33007 Oviedo, Spain
\\  $^{3}$ Instituto de F\'\i{sica} de Cantabria (CSIC-UC), Avda. los Castros  s/n, 39005 Santander, Spain
\\  $^{4}$ Departamento de F\'\i sica Moderna, Universidad de Cantabria, Avda. los Castros, s/n, 39005 Santander, Spain
\\  $^{5}$ Departamento de F\'\i{sica}, Universidad de Oviedo, Avda. Calvo Sotelo s/n, 33007 Oviedo,
Spain }

\begin{document}

\maketitle

\begin{abstract}
We propose a new detection technique in the plane based on an
isotropic wavelet family. This family is naturally constructed as an
extension of the Gaussian-Mexican Hat Wavelet pair and for that
reason we call it the Mexican Hat Wavelet Family (MHWF). We show the
performance of these wavelets when dealing with the detection of
extragalactic point sources in cosmic microwave background (CMB)
maps: a very important issue within the most general problem of the
component separation of the microwave sky. Specifically, flat
two-dimensional simulations of the microwave sky comprising all
astrophysical components plus instrumental noise have been analyzed
for the channels at 30, 44 and 70 GHz of the forthcoming ESA's {\it
Planck} mission Low Frequency Instrument (LFI). We adopt up-to-date
cosmological evolution models of extragalactic sources able to fit
well the new data on high-frequency radio surveys and we discuss our
current results on point source detection by comparing them with
those obtained using the Mexican Hat Wavelet (MHW) technique, which
has been already proven a suitable tool for detecting point sources.
By assuming a 5\% reliability level, the first new members of the
MHWF, at their ``optimal scale'', provide three point source
catalogues on half of the sky (at galactic latitude $|b|> 30^\circ$)
at 30, 44 and 70 GHz of 639, 387 and 340 extragalactic sources,
respectively. The corresponding flux detection limits are 0.38, 0.45
and 0.47 Jy . By using the same simulated sky patches and at the
same frequencies as before, the MHW at its optimal scale provides
543, 322 and 311 sources with flux detection limits of 0.44, 0.51
and 0.50 Jy, respectively (5\% reliability level). These results
show a clear improvement when we use the new members of the MHWF
and, in particular, the MHW2 with respect to the MHW.

\end{abstract}


\section{Introduction}
\footnotetext{E-mail: gnuevo@sissa.it}

The component separation of the microwave sky is one of the
relevant problems of cosmic microwave background (CMB) data
analysis. Before extracting the very important information encoded
in the CMB anisotropies (e.g. cosmological parameters, evolution
scenario, probability distribution of the primordial
perturbations) it is critical to separate the pure CMB signal from
other \emph{contaminant} emissions also present in the microwave
sky. These contaminants are known as foregrounds. Foregrounds are
usually divided in two categories: galactic emissions (dust,
free-free and synchrotron) and compact sources (galaxy clusters
and extragalactic sources). A detailed description of the expected
contamination level (both in terms of frequency and angular power
spectrum) for each foreground is discussed in \cite{te00}. The
emission due to extragalactic point sources (radio and infrared
galaxies seen as point-like objects due to their very small
projected angular size as compared to the typical experimental
resolutions) has very specific characteristics. Firstly, whereas
galactic foregrounds are highly anisotropic (most of their
emissions are concentrated inside the galactic plane),
extragalactic point sources are isotropically distributed in the
sky \citep{fra89,go05} and display an angular size determined by
the full width half maximum (FWHM) of the observational beam.

Secondly, whereas the energy spectrum of the galaxy clusters
emission at CMB frequencies is very well known and it is the same
for all the clusters, the energy spectrum of each point source is
given by the intrinsic characteristics of the physical processes
that occur inside each galaxy and can be quite different from
source to source.

Thirdly, whereas there are very good templates for the Galactic
components\footnote{See, e.g., the recently released Planck
``Galactic Reference Sky Model'' prepared by the Planck Working
Group 2, {\sl Component Separation}, and available at the web page
http://www.planck.fr/heading79.html}, not only about their spatial
distribution, but also about their frequency dependence, our
knowledge of the extragalactic point source emission at microwave
frequencies comes mostly from models based on observations at lower
and higher frequencies \citep{tof98,gui98,dole03,gra04}. This makes
the detection of as many extragalactic sources as possible in CMB
anisotropy maps a very important task by itself and not only in
terms of {\it map cleaning}.\footnote{Obviously, all-sky samples
made up of thousands of sources, hopefully available in the next
future with Planck, will be unique in providing knowledge on the
emission and cosmological evolution properties of the different
underlying source populations, in this almost unexplored frequency
domain\citep{dezo04}.}

Albeit very useful for studying the brightest sources in the CMB
sky, the very shallow survey (208 objects at fluxes $S\geq 1$ Jy)
provided by the NASA's WMAP satellite \citep{ben03a} does not allow
a detailed analysis of the statistical properties of the observed
source populations. The situation shall be partially clarified in
the near future, thanks to the ground-based very large sky surveys
(e.g., ATCA, Ryle Telescope) that have been already planned and
partially completed \citep{wal03,ric04} at $\nu\leq 30$ GHz. On the
other hand, there is no planned large area survey at frequencies
$\geq 30$ GHz, due to the limits of ground-based techniques.
However, the current lack of data should be overcome thanks to the
all-sky surveys at 9 frequencies (from 30 to 860 GHz) that should be
completed by the ESA {\it Planck} satellite \citep{man98,pud98}
during the year 2008. {\it Planck} surveys, intended to pick up from
several hundreds to thousands of extragalactic
sources\citep{dezo04}, shall be unique in this new observational
window: they will span a larger frequency interval and shall be also
much more sensitive than the WMAP ones. Moreover, the complementary
samples of point sources provided by the ESA {\it Herschel}
satellite, which will be launched jointly with {\it Planck} for
observing the sky at higher (mid- and far-IR) frequencies, should
help in clarifying the nature of source populations observed by
Planck HFI and of high-redshift sources, in particular.

For all the above reasons and for the fact that, at
mm/sub-millimeter frequencies, the emission of extragalactic
sources is at a minimum \citep{dezo99} -- making more difficult
their identification and separation from the other astrophysical
components -- it is of great interest the study and development of
new techniques for the detection of extragalactic point sources,
as well as the improvement of existing ones by reaching fainter
fluxes.

In particular, wavelet techniques have shown a very good performance
to solve this problem. Wavelets have been applied to many different
fields in physics during the last decade, e.g. geophysics, fluids
and astrophysics/cosmology.

A relevant approach was also introduced by \citet{mar80} considering
the continuous wavelet transform. In the latter case there is a
redundance of wavelet coefficients but analysis of signals and
images can also be performed. The basic idea, when we apply wavelets
in $R^N$, is the decomposition of a function $f(\vec{x})$ on a basis
that incorporates the local and scaling behavior of the function.
Therefore, apart from the domain of definition, the continuous
transform involves translations and dilations

\begin{equation}
w(\vec{b}, R) = \int d\vec{x}\,f(\vec{x})\Psi (\vec{x}; \vec{b},
R),
\end{equation}
\noindent with
\begin{equation}
 \Psi \Big(\vec{x}; \vec{b}, R\Big) \equiv \frac{1}{R^N}\psi
\Big(\frac{|\vec{x}-\vec{b}|}{R}\Big),
\end{equation}
\noindent where $\psi$ and $w$ are the mother wavelet and wavelet coefficient,
respectively, and $R$ is the scale (we assume an isotropic wavelet; see, e.g.,
\citet{san99}).

A particular and relevant case is the Mexican Hat Wavelet (MHW) defined on $R^2$

\begin{equation}
\psi (x)\propto (2 - x^2)e^{-x^2/2}, \ \ \ x\equiv |\vec{x}|.
\end{equation}

\noindent This wavelet and its extension to the sphere have been
extensively used in the literature to detect structure on a 2D
image: e.g. in astrophysics, for detecting point sources in CMB
maps \citep{cay00,vie01,vie03} and galaxy clusters in X-ray images
\citep{dam97}, by using the signal amplification when going from real
to wavelet space.

The Mexican Hat Wavelet is a very useful and powerful tool for point source
detection due to the following reasons:
\begin{itemize}
\item It has an analytical form that is very convenient when
making calculations and that allows us to implement fast algorithms.
\item It is well suited for the detection of Gaussian structures because it is obtained
by applying the Laplacian operator to the Gaussian function.
\item It amplifies the point sources with respect to the noise. Moreover, by changing the scale
of the Mexican Hat it is possible to control the amplification until an optimum value is achieved.
\item Besides, to obtain the optimal amplification it is not necessary to assume anything about the noise. In
Vielva et al. (2001), it was shown that the optimal scale can be
easily obtained by means of a simple procedure for any given image.
Therefore the Mexican Wavelet is a very robust tool.
\end{itemize}

We want to further develop the idea behind the Mexican Hat by
exploring a generalization of this wavelet that keeps these good
properties while improving the detection. In this paper we introduce
a natural generalization of the Gaussian-MHW pair on $R^2$ that
satisfies the properties mentioned above. This generalization will
allow us to improve the number of point source detections in CMB
maps, controlling the fraction of false detections (reliability). In
\S 2, we develop our technique and give some properties of the
Mexican Hat Wavelet Family (MHWF). In \S 3, we study first the case
of a point source embedded in white noise. Subsequently, we apply
the method to the problem of point source detection in CMB maps,
considering in particular the Low Frequency Instrument (LFI) of the
Planck mission. Realistic numerical simulations and a comparison
with standard (wavelet space) techniques are performed. Finally, the
main conclusions are drawn in \S 4.

\section{The Mexican Hat Wavelet Family}

  The MHW on the plane is obtained by applying the Laplacian operator
to the 2D Gaussian. If we apply the Laplacian to the MHW we obtain
a new wavelet and if we iterate the process we get a whole family
of wavelets: we call these wavelets the MHWF. We choose the
normalization so that the sum of the Fourier transforms of all
these wavelets and of the Gaussian filter is one. Therefore, any
member of the family can be written in Fourier space as
\begin{equation}
\hat\psi_{n}(k)=\frac{k^{2n}e^{-\frac{k^{2}}{2}}}{2^{n}n!}
\end{equation}
The new wavelet expression in real space is
\begin{equation}
\label{cc} \psi_{n}(x)=\frac{(-1)^n}{
2^{n}n!}\triangle^{n}{\varphi(x)}
\end{equation}
\noindent where $\varphi$ is the 2D Gaussian $
\varphi(x)=\displaystyle\frac{e^{-x^2/2}}{2\pi}$ and we apply the
Laplacian operator n times. Note that $\psi_{1}(x)$ is the standard
MHW.

The four first members of the MHWF are shown in real space in
Figure 1. Since the MHW has proven very useful for dealing with
point source detection, we will explore the performance of the
MHWF in this practical application.

Let us consider a field $f(\vec{x})$ on the plane $R^2$, where
$\vec{x}$ is an arbitrary point.
One can define the wavelet coefficient at scale $R$ at the point
$\vec{b}$ in the form given by equations (1), (2) with N=2.

\noindent The Fourier transform of $\psi_{n}(x)$ is

\begin{equation}
\label{cc}
 \hat\psi_{n} (k) = \int_0^{\infty}dx\,xJ_0(kx)\psi_{n} (x),
\end{equation}

\noindent where $\vec{k}$ is the wave number, $k\equiv |\vec{k}|$
and $J_0$ is the Bessel function of the first kind.

\noindent The wavelet coefficients  $w_n(\vec{b},R)$ for each
member of the MHWF can be obtained in the following form

\begin{equation}
\label{cc}
 w_n(\vec{b},R) = \int d\vec{k}e^{-i\vec{k}\cdot
\vec{b}}f(\vec{k})\hat\psi_{n} (kR),
\end{equation}

 \noindent this expression can be rewritten (we assume
the appropriate differential and boundary conditions for the field
$f$) as

\begin{equation}
\label{cc}
 w_n(\vec{b}, R)= \int
d\vec{x}\,[{\triangle^n}f(\vec{x})]\varphi\Big(\frac{|\vec{x}-\vec{b}|}{R}\Big).
\end{equation}

\noindent%
Hence, the wavelet coefficient at point $\vec{b}$ can be interpreted
as the filtering by a Gaussian window of the invariant
$(2n)^{\mathrm{th}}$ differences of the field $f$. We are then
decomposing the field(image) with this wavelet family and analyzing
it at different resolution levels.

\section{Source Detection with the MHWF}

The Mexican Hat Wavelet (MHW) was applied with great success to the
problem of detecting extragalactic sources in flat CMB maps
\citep{cay00,vie01}. The method was extended to the sphere
\citep{vie03} and applied successfully to detailed simulations of
maps at the Planck frequencies and with the characteristics of the
Planck experiment. In all these cases the MHW and its extension to
the sphere were used as a filter at different scales, enhancing the
signal to noise ratio for the sources and allowing us a very
efficient detection and a good determination of the source flux.
 In the following we will test the MHWF, as previously defined,
for source detection.

\subsection{Source detection in white noise maps}

A basic point when we use a filter for source detection is how the
signal to noise ratio of the sources in the filtered map is
increased with respect to that of the original map; this effect is
called amplification. In our first example, we consider a point
source with a Gaussian profile embedded in white noise (i.e. a
homogeneous and isotropic random field with a constant power
spectrum). We consider a flat pixelized image consisting of a
filtered source plus white noise added at any pixel and then we
calculate the amplification produced by the MHWF in this particular
case. Let us first calculate analytically, see eq. (7), the wavelet
coefficients obtained by filtering with any member of the MHWF  a
source of temperature $ T(x)=T_{0} e^{-\frac{x^{2}}{2\gamma^{2}}}$,
where the source of temperature $T_{0}$ has been previously
convolved with a Gaussian beam of dispersion $\gamma$. We obtain the
following coefficients at $\vec{b}=0$, (the point of maximum
temperature of the source)

\begin{equation}
\label{cc}
 w_{n}=\displaystyle\frac{T_0 \beta^{2n}}{(1+\beta^2)^{n+1}}
\end{equation}

\noindent where $\beta=\frac{R}{\gamma}$ and R is the wavelet
scale. Now we will calculate the rms deviation $\sigma_{w_{n}}$ of
the background (white noise) filtered with a wavelet of scale R as
a function of the rms deviation $\sigma$ of the background.

\begin{equation}
\label{cc}
 \sigma_{w_{n}}=\displaystyle\frac{\sigma l_{p}
\sqrt{(2n)!}}{R \pi 2^{n} n!}
\end{equation}

\noindent where $l_{p}$ is the pixel size. The amplification
$\lambda_n$ is defined as
\begin{equation}
\label{cc}
\lambda_n=\displaystyle\frac{\displaystyle{w_{n}/\sigma_{w_{n}}}}{\displaystyle{T_0/\sigma}}
\end{equation}

Then, the amplification for the white noise case can be written as

\begin{equation}
\label{cc}
 \lambda_{n}=\displaystyle\frac{\beta^{2n+1} \gamma \pi 2^n n!}{(1+\beta^2)^{n+1}
 l_{p}
 \sqrt{(2n)!}}
\end{equation}

We can obtain the scale of maximal amplification for each wavelet,
deriving with respect to $\beta$ and equating to zero. We
obtain $\beta_{max}=\sqrt{2n+1}$, so that the optimal scale
depends on n and is $R_{max}=\sqrt{2n+1} \gamma$. The maximum
amplification only depends on $n$ and can be written

\begin{equation}
\label{cc}
 \lambda_{max}=\displaystyle\frac{(2n+1)^{(n+\frac{1}{2})} n!
 \pi \gamma}{2(n+1)^{n+1} \sqrt{(2n)!} l_{p}}
\end{equation}

If we calculate $\lambda_{max}$ according to this formula, we obtain
a maximum value for $n=0$, a Gaussian filter with $R=\gamma$. So,
the optimal amplification is reached when we filter again with a
Gaussian beam of the same dispersion as the original one. The
maximum amplification for other members of the MHWF decreases slowly
as the index n increases.

\subsection{Source detection in CMB maps}

We have carried out a simple calculation for the case of a point
source embedded in a white noise background. However, we are more
interested in the realistic conditions of a CMB experiment in which
the source is embedded in the cosmic microwave background, the
Galactic foregrounds and the detector Gaussian noise. A way to
analyze the amplification given by (11) is to carry out simulations
of the CMB maps with all the components and measure the rms
deviations $\sigma_{w_n}$ and $\sigma$. The expression for $w_n$ is
the one given by (9).

For detecting point sources and for calculating the amplification
factor of a source embedded in a realistic CMB map, we have carried
out simulations of CMB 2D maps of $12.8\times12.8$ square degrees,
generated with the cosmological parameters of the standard model
\citep{spe03}. We have then added the relevant Galactic foregrounds
(free-free, synchrotron and dust emission) by using the Planck
Galactic Reference Sky Model, provided by the members of the Planck
Working Group 2.

  In order to give the (hopefully) most realistic numbers of detected
extragalactic point sources, we have adopted the cosmological
evolution model for sources recently presented by \citet{dezo05} for
predicting the numbers of galaxies in our simulations. This new
model, which takes into account the new data coming from high
frequency ($\sim 20-30$ GHz) radio surveys of extragalactic sources
and which discusses all source populations that can contribute to
number counts at LFI frequencies, has proven capable of giving a
better fit than before to all the currently published source number
counts and related statistics at $\nu\leq 30$ GHz coming from
different surveys \citep{wal03,ben03b,mas03,ric04,cle05}.

Since we are specially interested in applying our new method to the
maps which will be provided by the {\it Planck} mission in the next
future, and given that we can be very confident in the input source
model counts, we have considered the specific conditions of the LFI
{\it Planck} channels, which operate at 30, 44 and 70
GHz.\footnote{The characteristics of the LFI channels, relevant for
our purposes, are: a) pixel sizes, $6'$, $6'$ and $3'$ at 30, 44 and
70 GHz, respectively; b) Full Width Half Maximum (FWHM) of the
circular gaussian beams, $33'$, $24'$ and $14'$, respectively; c)
thermal (uniform) noises, $\sigma=2\times10^{-6}$,
$\sigma=2.7\times10^{-6}$ and $\sigma=4.7\times10^{-6}$,
respectively, in a square whose side is the FWHM extent of the beam.
In all the cases, we have used the estimated instrument performance
goals available at the web site: http://sci.esa.int/planck.}

According to formula (9) we can calculate the temperature of the
point source at its central pixel in the filtered maps for each
wavelet and at any scale, given the original temperature. We find
complete agreement between the values obtained from the formulas and
the results coming from our simulations for all the wavelets and
scales involved. Our next step is to calculate the amplification
for different members of the MHWF and for different values of the
wavelet scale, R; our goal is to obtain the {\it optimal scales}
for the different wavelets used and to compare them for source
detection.  At 30 GHz, the maximum amplification for the MHW is
$2.21$. This amplification is reached at the optimal scale
$R=9.1'$. For the other members of the family, MHW2, MHW3 and
MHW4, the amplifications are: 2.68, 2.87, 2.94 and are obtained at
the optimal scales $R=13.5', 17.4'$ and $20.5'$, respectively.

At 44 GHz the maximum amplification for the MHW is $2.3$ and it is
obtained at the scale $R=7.3'$. For the members MHW2, MHW3 and
MHW4 the amplifications are: 2.74, 2.83, 2.80 and correspond to
the scales $R=11.2',14.1'$ and $16.5'$ respectively. In Figure 2
we show, as an example, the amplification, $\lambda$, obtained at
different scales with the four first members of the MHWF at 44
GHz. At 70 GHz and for the MHW, MHW2, MHW3 and MHW4 we find
$\lambda=$2.59, 2.79, 2.76, 2.69, and these amplifications are
obtained at the scales $R=5.3', 7.6', 9.4'$ and $10.9'$
respectively. All the tests done with members of the MHWF defined
by higher $n$ (see Eq. 5) show a lesser amplification at each of the
three frequencies and this also happens when a Gaussian filter is
used. For this reason we limited our study of the capability of
source detection of the MHWF  to the first four members of the
family.

We have then used the MHW2, MHW3 and MHW4 at the optimal scales to
detect point sources in the maps at 30, 44 and 70 GHz, comparing the
results with those obtained with the MHW. Our detection method is
quite simple, we carry out enough simulations ($10\times 126 $) of
2D sky patches to cover ten times half the sky at $|b|>30^\circ$. As
previously stated, the simulations include point sources, Galactic
foregrounds, CMB and the detector noise.

In Figure 3 (top left panel) we show one of the simulated CMB maps
we analyzed at 44 GHz and the corresponding map of extragalactic
point sources only (top right). The same map, filtered with the MHW
and with the MHW2 at their optimal scales, is shown in the
central-left and central-right panels, respectively. In the
bottom-left and bottom-right panels we show the same original
simulated map filtered with the MHW3 and the MHW4, respectively. In
all the panels the brightest (detected) sources are shown as red
spots in the filtered maps (given the adopted color scale of the
maps; see caption); since we have simulated sky patches of
$12.8\times12.8$ square degrees, only a few (very bright) sources
are detected in each sky patch. Thus, from only one map, it is not
possible to tell visually apart the differences in the number of
detections after filtering with one specific member of the family or
with another one. However, the map filtered with the MHW
(central-left panel) is clearly noisier than the other maps filtered
with other members of the Family. This is due to the fact that the
variance of this map is $\simeq 10\%$ higher than in the other
cases.

 After filtering each simulation with the corresponding wavelet at the optimal scale,
we use equation (9) to calculate the flux at any pixel. First, we
select those pixels corresponding to maxima above 300 mJy (since for
lower flux detection limits the fraction of spurious sources, as
defined below, is very high). Then we compare these maxima with the
fluxes of the sources in the simulations (beam filtered). We
consider a detection as real when the following criteria are
satisfied: a) the distance between each maximum and the
corresponding simulated source is $\leq FWHM/2$, b) the relative
difference between the estimated and the real flux is less than
$100\%$. The maxima (above 300 mJy) which do not fulfil these
criteria are considered as spurious sources.

The average number of sources correctly detected in half of the sky
(see Tables 1, 2 and 3) with the MHW, MHW2, MHW3 and MHW4,
respectively, are: 873, 850, 846 and 845 at 30 GHz; 695, 673, 671
and 674 at 44 GHz; 634, 622, 623 and 627 at 70 GHz. The average
number of spurious sources with the MHW, MHW2, MHW3 and MHW4,
respectively, are: 1230, 431, 409 and 526 at 30 GHz; 4538, 2008,
1988 and 2366 at 44 GHz; 8082, 4980, 5236 and 6254 at 70 GHz. Since
the number of spurious sources for the assumed flux limit (300 mJy)
is very high, it seems more appropriate to impose a level of
reliability (fraction of spurious sources), e.g. $5\%$. With this
reliability level the number of real detections for the MHW, MHW2,
MHW3 and MHW4 are: 543, 639, 583 and 418 at 30 GHz;  322, 387, 366
and 275 at 44 GHz;  311, 340, 331 and 288 at 70 GHz.

In Table 1, these results at 30, 44, 70 GHz are shown together with
other relevant information about the detection process: the flux
limit for which the reliability is $5\%$, the flux over which the
source catalogue is $95\%$ complete, the number of sources detected
in the $95\%$ completeness catalogue and the average of the absolute
value of the flux determination relative error. In the table, we
write the average results and the rms deviations (from ten half
skies).

We want to remark the following results: 1) the average number of
detections above 300 mJy is a little higher for the MHW than for the
other members of the MHWF, but the number of spurious sources is
much higher; 2) the number of detected sources with a $5\%$
reliability is higher for the MHW2 than for the other members; when
we compare with the MHW, we see a clear increase in the real
detections, this increase is $\simeq 18\% $ (30 GHz), $\simeq 20\% $
(44 GHz), $\simeq 9\% $ (70 GHz); the perfomance of the MHW4 is the
worst in this respect; 3) we also note that the average error in the
flux determination is always below $\simeq 25\% $, being lower for
the new wavelets than for the MHW.

According to these results, the most efficient strategy is to study
all these wavelets, MHW$n$ with $n\leq 4$, and choose the one that better fulfils the detection criterion.
Of course, this conclusion can be applied only to the kind of maps used here; in other Planck
channels, for example, the best results may be obtained with other
MHW$n$s. In general, the use of the MHWF allow us to determine and
(then) apply the most efficient wavelet filter, of the same family,
for detecting point sources embedded in realistic CMB maps. We want
to point out that the proposed method could be applied to the real
maps provided by the Planck satellite, since it is easy to
calculate, in each particular case, the amplification and the
optimal scales for source detection.

In Figure 4 (top left panel) we plot the percentage of spurious
sources against the flux detection limit for the different wavelets
at 30 GHz. In the middle left panel, we show the same results at 44
GHz and in the bottom left panel at 70 GHz. As it can be seen, this
percentage is higher for the MHW and the MHW4 than for the other
wavelets. In the same figure, we plot the absolute error in the flux
determination against the flux detection limit at 30 GHz (top
central panel), 44 GHz (middle central panel) and 70 GHz (bottom
central panel). As it is clear from the Figure, this error is lower
for the new members than for the standard MHW. Finally, in the right
panels we plot the relative error in the flux determination for the
three frequencies; the error percentage is also lower for the MHW2
and MHW3 than for the other wavelets.

We have analysed (realistic) simulated CMB maps at three
frequencies, corresponding to the LFI channels of the {\it Planck}
mission, but the present results encourage us to test the proposed
method at other CMB frequencies, in forthcoming papers. Moreover,
the method could be extended to the sphere, just by applying the
stereographic projection \citep{cay01,mar02} to the MHWF and then
by using it to analyze spherical CMB maps \citep{vie03}.

\section{Conclusions}

We have considered a natural generalization of the MHW on the
plane, $R^2$, based on the iterative aplication of the Laplacian
operator to the MHW. We have called this group of wavelets the
Mexican Hat Wavelet Family (MHWF).

The MHWF can therefore be applied to the analysis of flat
two-dimensional images and, in particular, to current and
forthcoming CMB  maps provided by the plethora of the existing (and
planned) experiments. Our main goal is to test this wavelet family
in one typical application: point source detection in CMB anisotropy
maps and to compare our results with those obtained by the use of
the MHW. The detection of as many extragalactic sources as possible
in CMB maps is an important issue {\it per se}, and not only for the
purpose of map cleaning, for it allows the construction of
catalogues of point source populations at frequencies where there
are only few and sparse data available at the present time.

The main reason why wavelets perform well when detecting point
sources is that they amplify the ratio between the point source
flux density and the dispersion of the background. For a simple
background, such as white noise, this amplification can be
calculated easily, see eqs. (12) and (13). In this very simple case
the maximum amplification is obtained with the Gaussian filter.

 However, since we want to propose a detection method that could be
applied to real CMB anisotropy maps, we have carried out realistic
simulations of the CMB sky including primordial CMB anisotropies,
point sources, instrumental (thermal Gaussian) noise and all the
relevant foregrounds. To test the method, we have chosen the three
channels of the Planck LFI instrument, 30, 44 and 70 GHz (using the
characteristics currently established for them). This choice is also
due to the fact that, in this frequency interval, it is currently
possible to simulate a very realistic number of bright extragalactic
point sources in the sky, by exploiting the cosmological evolution
model of \citet{dezo05}. In all the cases we have considered
$10\times126$ maps of $12.8\times12.8$ square degrees for covering
ten times half the sphere at high Galactic latitude. After filtering
with the MHWF, we select the maxima in such maps as possible
candidates and compare them with the maxima in the beam filtered
source maps.

We consider a detection as real if the two following criteria are
satisfied a) the distance between the maxima and the corresponding
sources is $\leq FWHM/2$, b) the relative difference between the
estimated and the real flux is less than $100\%$. In this way, we
construct a source catalogue above  $300$ mJy  of real detections.
The other maxima are considered as spurious sources. From this
catalogue we can draw the following main conclusions:

a) We have compared the performance of the new members of the MHWF
for point source detection with that of the MHW alone, at their
optimal scale \citep{cay00,vie01}. Our results clearly show that,
although the number of detections is similar, the {\it number of
spurious sources is much lower} ($\leq 50\%$) for the new wavelets
than for the MHW.

b) By assuming a $5\%$ reliability, the {\it number of detected
sources} obtained by filtering with the new wavelets and, in
particular, with the MHW2 is {\it higher} than the number obtained
with the original MHW (see Section 3.2). We detect (639/543) sources
(MHW2/MHW) at 30 GHz, (387/322) sources (MHW2/MHW) at 44 GHz and
(340/311) sources (MHW2/MHW) at 70 GHz. The $ 5\% $ reliability flux
limit is (380/440) mJy at 30 GHz, (450/510) mJy at 44 GHz and
(470/500) mJy at 70 GHz.

c) The average of the absolute value of the relative error ($|\Delta
S|/S$) is {\it lower} for the new members of the MHWF than for the
MHW. This average value is (18.1/21.9)\%  at 30 GHz, (21.6/25.5) \%
at 44 GHz and (21.4/23.5) \% at 70 GHz.

In general, we can see a clear improvement in the number of
detections and in the flux estimation when we apply the new
wavelets. Further information can be seen in Table 1. In Figure 4 it
can also be seen the percentage of spurious sources and the average
absolute and relative errors in the flux determination plotted
against the flux limit for the different wavelets and frequencies.

 The MHWF could be extended to the
sphere in a straightforward way, by using a stereographic
projection, allowing us to analyze all-sky CMB maps. CMB
experiments other than \textit{Planck} could also benefit from the
source amplification performed by the MHWF.

\section*{Acknowledgments}

We acknowledge partial financial support from the Spanish Ministry
of Education (MEC) under projects ESP2002--04141--C03--01 and
ESP2004--07067--C03--01. JGN and MLC acknowledge a FPU and FPI
fellowship of the Spanish Ministry of Education and Science (MEC) ,
respectively. DH acknowledges a Juan de la Cierva contract of the
MEC. We thank G. De Zotti for having kindly provided us with the
source number counts foreseen by the De Zotti et al. (2005)
cosmological evolution model at the LFI frequencies. We acknowledge
the use of the Planck Reference Sky, prepared by the members of the
Planck Working Group 2, and available at
www.planck.fr/heading79.html. We also thank E.
Mart\'\i{nez}-Gonz\'alez, R.B. Barreiro, A. Aliaga \& G.
Gonz\'alez-Nuevo for useful discussions.



\begin{table*}
  \centering
  \begin{tabular}{cccccccc}
  \textbf{30 GHz}& \textbf{N$_d$}& \textbf{N$_{sp}$} & \textbf{S$_{95}$(Jy)}& \textbf{N$_{95}$}& \textbf{S$_{5}$(Jy)}& \textbf{N$_{5}$}& \textbf{$|\Delta S|/S$}\\
  \hline\hline
    \textbf{MHW} & 873(30)  &1230(39)  &0.33(0.01)  &637(26)  &0.44(0.01)  &543(26) &21.9(0.9) \\
    \textbf{MHW2} &850(27)  &431(32)  &0.34(0.02)  &632(27)  &0.38(0.01)  &639(31)  &18.6(0.5) \\
    \textbf{MHW3} &846(34)  &409(36)  &0.34(0.01)  &628(24)  &0.41(0.02)  &583(42)  &18.1(0.4) \\
    \textbf{MHW4} &845(35)  &526(39)  &0.34(0.01)  &626(28)  &0.51(0.04)  &418(63)  &18.4(0.5) \\
  \hline\hline
\textbf{44 GHz} & \textbf{N$_d$}& \textbf{N$_{sp}$} & \textbf{S$_{95}$(Jy)}& \textbf{N$_{95}$}& \textbf{S$_{5}$(Jy)}& \textbf{N$_{5}$}& \textbf{$|\Delta S|/S$}\\
  \hline\hline
    \textbf{MHW} & 695(28) &4538(66)  &0.35(0.01)  &459(32)  &0.51(0.01)  &322(22) &25.5(0.6) \\
    \textbf{MHW2} &673(26)  &2008(43)  &0.35(0.01)  &462(24)  &0.45(0.01)  &387(21)  &21.7(0.6) \\
    \textbf{MHW3} &671(26)  &1988(39)  &0.34(0.01)  &467(31)  &0.46(0.02)  &366(37)  &21.6(0.7) \\
    \textbf{MHW4} &674(28)  &2366(43)  &0.35(0.01)  &466(33)  &0.56(0.06)  &275(54)  &22.1(0.7) \\
  \hline\hline
\textbf{70 GHz}& \textbf{N$_d$}& \textbf{N$_{sp}$} & \textbf{S$_{95}$(Jy)}& \textbf{N$_{95}$}& \textbf{S$_{5}$(Jy)}& \textbf{N$_{5}$}& \textbf{$|\Delta S|/S$}\\
  \hline\hline
    \textbf{MHW}&634(24) &8082(72)  &0.33(0.01)  &452(23)  &0.50(0.01)  &311(24)  &23.5(0.9)  \\
    \textbf{MHW2}&622(30) &4980(81)  &0.34(0.01)  &447(31)  &0.47(0.01)  &340(31)  &21.4(0.8)   \\
    \textbf{MHW3}&623(27) &5236(108)  &0.34(0.01)  &443(34)  &0.47(0.01)  &331(30)  &21.7(0.9)  \\
    \textbf{MHW4}&627(26) &6254(105)  &0.34(0.01)  &444(36)  &0.53(0.03)  &288(38)  &22.3(0.9)    \\
  \hline\hline
\end{tabular}
  \caption{Point source detections at 30, 44 and 70 GHz in half of the sky, $2\pi$ sr, above $300$ mJy , for several members of the
  MHWF. In each column we indicate the average and the r.m.s. deviation (in curl brackets) of some relevant quantities calculated by the
  simulations on sky patches corresponding to ten half skies.
  In the first column, we show the number of detections, $N_{d}$, in the second one the number
  of spurious sources, $N_{sp}$, and in the third one we list the fluxes at which the recovered source
  catalogue are complete at the $95\%$ level. The number of detected sources brighter than these fluxes are shown
  in the fourth column. In the fifth and sixth column, the fluxes above which
  the catalogues have a $5\%$ reliability and the corresponding numbers of
  detected sources are shown, respectively. Finally, we show in the last column the average of the absolute
  value of the relative error (in percentage) in the flux determination.
  }\label{table1}
\end{table*}

\clearpage

\begin{figure*}
 \epsffile{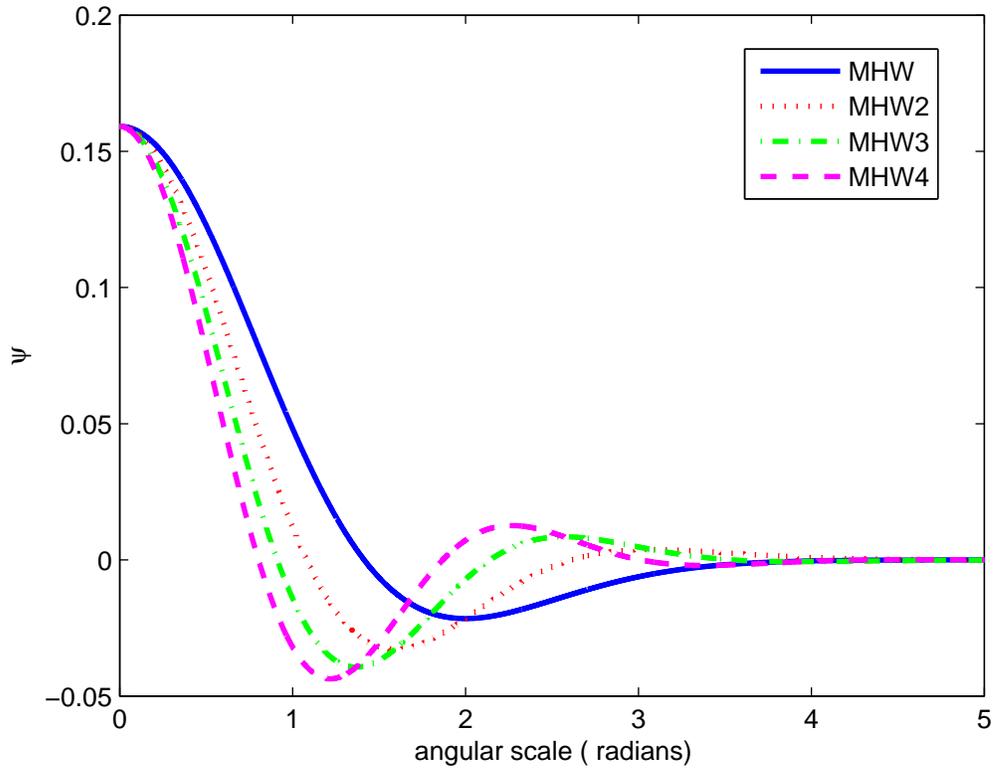} \caption{\label{fig1}{\small
 Radial profile of the Mexican Hat Wavelet (solid line), MHW2 (dotted line), MHW3 (dash-dotted line), MHW4 (dashed line).}}
\end{figure*}

\clearpage

\begin{figure*}
\epsffile{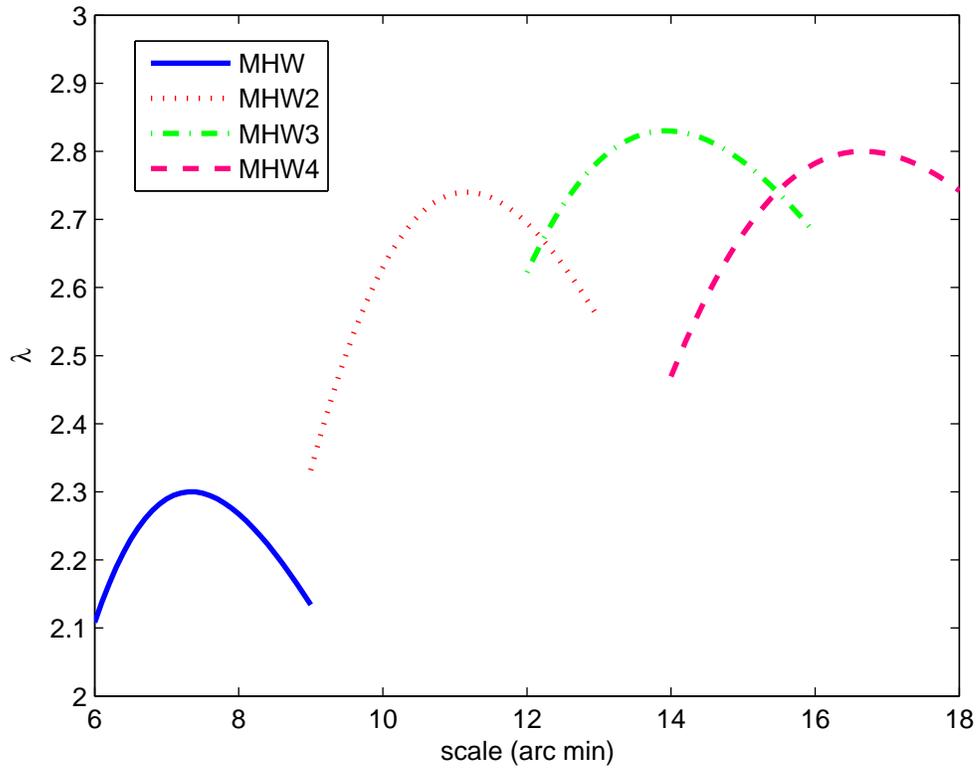} \caption{\label{fig2}{\small Amplification
$\lambda$ of a point source in 44 GHz CMB maps for the MHW (solid
line), MHW2 (dotted line), MHW3 (dash-dotted line) and MHW4 (dashed
line) as a function of the scale in arcminutes .}}
\end{figure*}

\clearpage

\begin{figure*}
\epsfxsize=200mm
 \epsffile{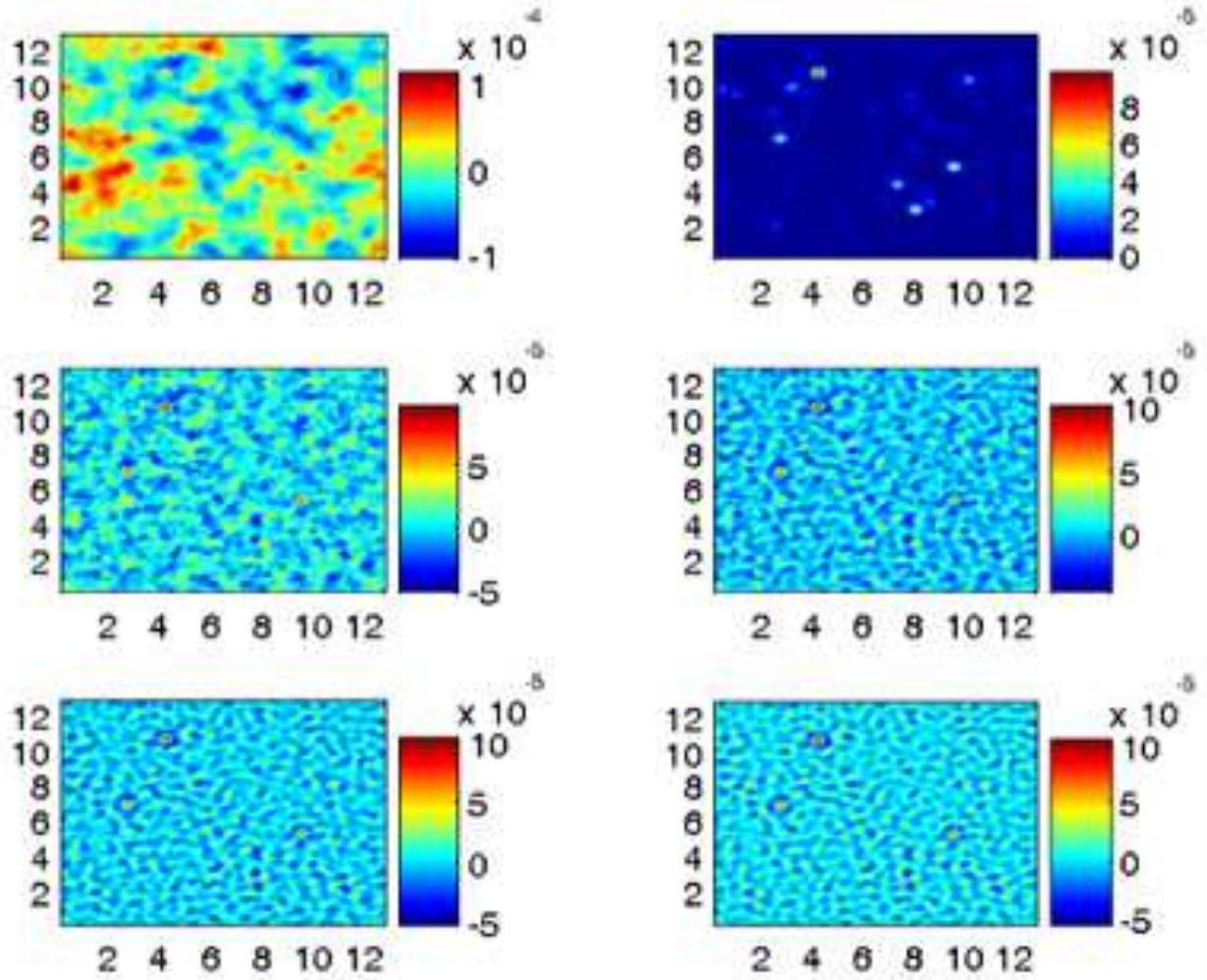}
 \caption{\label{fig3}{\small  44 GHz maps.
Top-left panel: CMB+Sources+Noise+Foregrounds map (total map);
Top-right: the corresponding map of simulated extragalactic point
sources only.
 From central-left to bottom-right, the panels show the total map filtered with the MHW, MHW2, MHW3, MHW4  at the
 corresponding optimal scale, respectively. The color scale is in
 $\delta T/T $ units. The angular scale is in degrees}}
\end{figure*}

\clearpage

\begin{figure*}
\epsfxsize=200mm
 \epsffile{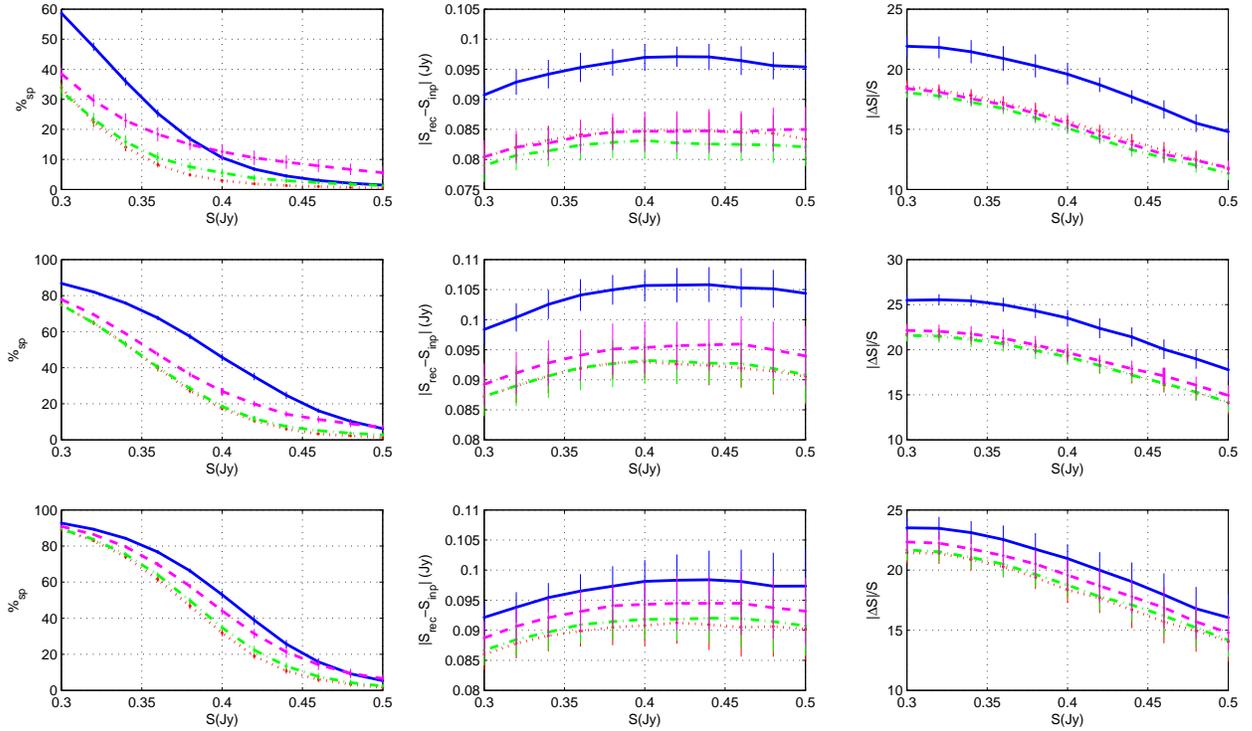} \caption{\label{fig4}{\small
Top left panel: the percentage of spurious sources at 30 GHz is
 plotted against the flux detection limit.
Top central panel: the average of the absolute value of the error in
the flux determination at 30 GHz is plotted against the flux limit.
Top right panel: the average of the absolute value of the relative
error (percentage) in the flux determination at 30 GHz is plotted
against the flux limit. In each panel the average of 10 half skies
together with the rms error bars are shown for the MHW (solid line),
MHW2 (dotted line), MHW3 (dash-dotted line) and MHW4 (dashed line).
(Middle/bottom) left panel: the same plot as in the top left panel
at (44/70) GHz. (Middle/bottom) central panel: the same plot as in
the top central panel at (44/70) GHz . (Middle/bottom) right panel:
the same plot as in the top right panel at (44/70) GHz .}}
\end{figure*}

\end{document}